\documentclass[aps,pra,twocolumn]{revtex4-1}
\usepackage{graphicx}
\usepackage{bm}
\usepackage{amsmath}
\usepackage[colorlinks=true,allcolors=blue]{hyperref}

\usepackage{float}

\graphicspath{{./images/}}

\providecommand{\abs}[1]{\left\lvert#1\right\rvert}

\newlength{\wdth}

\providecommand{\ket}[1]{\lvert #1 \rangle}
\providecommand{\braket}[2]{\langle #1 \rvert #2 \rangle}
\providecommand{\ketbra}[2]{\lvert  #1\rangle \langle #2 \rvert}
\providecommand{\be}{\begin{equation}}
\providecommand{\ee}{\end{equation}}
\providecommand{\ba}{\begin{eqnarray}}
\providecommand{\ea}{\end{eqnarray}}

\usepackage{mathbbol}
\usepackage[makeroom]{cancel}

\newcommand{\tx}[1]{\text{#1}}
\newcommand{\tb}[1]{\textbf{#1}}

\newcommand{\td}[1]{\widetilde{#1}}
\newcommand{\dagg}[1]{#1^\dagger}
\newcommand{\adag}{a^\dagger}
\newcommand{\sigx}{\sigma_x}

\newcommand{\sigz}{\sigma_z}
\newcommand{\sigp}{\sigma_+}
\newcommand{\sigm}{\sigma_-}
\newcommand{\D}{\mathcal{D}}
\newcommand{\Ddag}{\mathcal{D}^\dagger}

\newcommand{\wt}{\td \omega}

\usepackage{xcolor}

\begin{document}
\title{Trapped ions beyond carrier and sideband interactions}
\author{T. Tassis}
\affiliation{Centro de Ci\^encias Naturais e Humanas, Universidade Federal do ABC (UFABC), Santo Andr\'e, SP, 09210-580, Brazil}
\author{F. L. Semi\~ao}
\affiliation{Centro de Ci\^encias Naturais e Humanas,  Universidade Federal do ABC (UFABC), Santo Andr\'e, SP, 09210-580, Brazil}
\begin{abstract}
Trapped ions driven by electromagnetic radiation constitute one of the most developed quantum technologies to date. The scenarios range from proof-of-principle experiments to on-chip integration for quantum information units. In most cases, these systems have operated in a regime where the magnitude of the ion-radiation coupling constant is much smaller than the trap and electronic transition frequencies. This regime allows the use of simple effective Hamiltonians based on the validity of the rotating wave approximation. However, novel trap and cavity designs now permit regimes in which the trap frequency and the ion-radiation coupling constant are commensurate. This opens up new venues for faster quantum gates and state transfers from the ion to a photon, and other quantum operations. From the theoretical side, however, there is not yet much known in terms of models and applications that go beyond the weak driving scenario. In this work, we will present two main results in the scenario of stronger drivings. First, we revisit a known protocol to reconstruct the motional Wigner function and expand it to stronger driving lasers. This extension is not trivial because the original protocol makes use of effective Hamiltonians valid only for weak drivings. The use of stronger fields or faster operations is desirable since experimental reconstruction methods of that kind are usually hindered by decoherence. We then present a model that allows the analytical treatment of stronger drivings and that works well for non-resonant interactions, which are generally out of the reach of the previous models.
\end{abstract}
\pacs{}
\vskip2pc

\maketitle

\section{Introduction}
Trapped ions manipulated by classical and quantum radiation fields have now become flexible platforms to test quantum protocols  \cite{review,wine,matthias}. These include the generation and detection of non-classical states \cite{wine,gato,thermali,Hastrup} and the engineering of quantum operations \cite{cz,qre,Leung,qreo}, just to name a few examples. Typically, the experiments have been conducted in a regime where the applied radiation fields weakly interact with the electric dipole referred to the internal degrees of freedom of the ion. If $\Omega$ denotes the ion-radiation coupling constant, \emph{i.e.}, the Rabi frequency, and $\nu$ ($\omega_0$) denotes the trap (electronic transition) frequency, that regime is characterized by $\Omega\ll\nu,\omega_0$. This greatly facilitates the theoretical description of the system because the coupling to the external radiation fields can be regarded as a small perturbation. In this way, the rotating wave approximation (RWA) can be safely applied \cite{rwaperturb} what results in the well known carrier and sideband Hamiltonians \cite{review}. These are effective Hamiltonians which give accurate results only when $\Omega\ll\nu,\omega_0$.

In spite of the remarkable success of that regime, there are many reasons to pursuit accurate models to describe stronger couplings to the external fields. We would like to mention two of them. Firstly, by strengthen the coupling constant to $\Omega\sim\nu$, one could aim at designing fast quantum operations driven by the external laser. Given that  decoherence times often impose limits on the scaling of the protocols, it might be useful to operate at the highest possible speeds. Indeed, this is the main motivation for the proposal reported in \cite{Jonathan}.  There, the authors make use of a particular instance of the transformation originally reported in \cite{Cessa}. As we will see later on in this article, that transformation works well for $\Omega\sim\nu$ but it tends to lose accuracy as one moves the system out of exact resonance as given by the relation $\omega_L=\omega_0$, where $\omega_L$ is the laser frequency. One of our main results is precisely the presentation of a model which works well for $\Omega\sim\nu$ and is not restricted to exact resonance. Secondly, the exploration of new regimes usually favors the proposal of new applications. A representative example is the Rabi Hamiltonian as the non-RWA version of the well known Jaynes-Cummings model (JCM). While it is absolutely true that the JCM is very successful, its validity is only guaranteed for weak atom-radiation coupling constants. By allowing the use of stronger fields, which make the RWA unfeasible, the Rabi model allows, for example, the description of quantum phase transitions \cite{rabi0,rabi1}, something that is out of reach of the JCM.  

In the first part of this work, we will revisit one of the most relevant techniques in the toolbox used to measure and control trapped ions, which is the Wigner function reconstruction of the ionic motional state \cite{wigner1,wigner2}. The known protocols are based on effective Hamiltonians valid only for weak interactions with the external lasers, \emph{i.e.}, $\Omega\ll\nu$. Here, we generalize it to the use of stronger ion-radiation couplings, with $\Omega\sim\nu$. In order to illustrate the importance of using stronger couplings, we add some effective decoherence during the reconstruction protocol and compare the performance of our protocol with previous ones \cite{wigner1,wigner2}. In the second part of this work, we take a step further and show that the combination of unitary transformations \cite{Cessa,Jonathan,interfer,cessa-review}, RWA and diagonalization allows one to obtain a new effective Hamiltonian that is exactly diagonalizable. This Hamiltonian significantly enhance the range of parameters under which previous models produce accurate results \cite{Jonathan,Cessa}, specially in the non-resonant case $\omega_L\neq\omega_0$. 

This article is organized as follows. In Sec. \ref{basic}, we provide a short introduction to the basic concepts concerning the interaction of trapped ions and lasers. In particular, we briefly review the preceding models treating the regime $\Omega\sim\nu$. As said before, these models are accurate only when the laser is resonant with the internal levels of the ion. In Sec. \ref{wigner}, we present a new protocol to reconstruct the Wigner function in that resonant regime, taking advantage of  having faster operations to combat the detrimental effects of decoherence. In Sec. \ref{new}, we present a new model for the regime $\Omega\sim\nu$ which provides analytical results accurate even in the non-resonant case. Finally, we summarize and make final remarks in Sec. \ref{remarks}.

\section{Interaction of a trapped ion with a laser}\label{basic}
In this section, we provide a short review of the usual models for the interaction of a two-level trapped ion with a laser. As we will see, the main ingredient in those models is the assumption of a weak ion-laser interaction in order to obtain effective RWA Hamiltonians that can be analytically diagonalized. We will also review the few approaches found in the literature for stronger interactions, which is the regime treated in this article. 

For practical considerations, one usually models this system as a two-level ion whose center of mass is subjected to a harmonic potential. An external classical laser is then used to couple the internal and external degrees of freedom of the ion. Following a well-known microscopic derivation  \cite{review}, one is led to write the system Hamiltonian as
\begin{equation} \label{eq:hamilt-complete}
    H = H_0 + V,
\end{equation}
where 
\begin{equation} \label{eq:hamilt-0}
    H_0 = \nu \adag a + \frac{\delta}{2} \sigz
\end{equation}
is the free Hamiltonian of the trapped ion, with $\nu$ being the trap secular frequency and $\delta \equiv \omega_0 - \omega_L$, the laser-ion detuning.
Also,
\begin{equation} \label{eq:hamilt-V}
V=\frac{\Omega}{2} \biggl( \sigp \, \D(i \eta) + \sigm \, \Ddag(i \eta) \biggr),
\end{equation}
where $\eta$ is the Lamb-Dicke parameter \cite{review}, $\D(\alpha) \equiv e^{\alpha \adag - \alpha^* a}$ is the displacement operator and, as said before, $\Omega$ is the ion-laser coupling constant, also known as as the Rabi frequency. The coupling term in Eq. (\ref{eq:hamilt-V}) is seen as the result of momentum conservation and is the cause of entanglement between internal and external degrees of freedom. By keeping the relative angle between the laser wave vector and the trap axis upon which the ion is oscillating fixed, the Rabi frequency can be enhanced by increasing the laser power \cite{review,wine}.


The next step is usually the assumption of $\eta\ll 1$. This allows us to expand the exponentials  in Eq. (\ref{eq:hamilt-V}) up to $\mathcal{O}(\eta)$, leading to
\begin{equation} \label{eq:hamilt-V-lb}
    V \approx \frac{\Omega}{2} \sigx + \frac{i \eta \Omega}{2} (\sigp - \sigm) (a + \adag).
\end{equation}
 Typically, $\eta\lesssim 10^{-1}$ or less is allowed within this approximation which became known as the Lamb-Dicke regime \cite{wine}.

By tuning the laser frequency to satisfy $\delta = \nu$ and neglecting fast-oscillating terms, we arrive at the so-called first red sideband Hamiltonian
\begin{equation} \label{eq:hamilt-rsb}
    H_\tx{RSB} \approx \nu \adag a + \frac{\delta}{2} \sigz + \frac{i \eta \Omega}{2} \left( \sigp a - \sigm \adag \right),
\end{equation}
which is analogous to the JCM.
The validity of this RWA is heavily dependent on the magnitude of the coupling constants accompanying the neglected terms relative to the other frequencies involved, $\nu$ and $\delta$.
 Given that $\delta$ is usually of the order of $\nu$, the validity of that RWA requires $\Omega \ll \nu$.
This leads to a limitation in the speed of operations when considering this effective Hamiltonian, since the Rabi frequency dictates the rate of transitions in the JCM.

In order to access faster regimes, it is useful to have models valid for $\Omega \sim \nu$. This can be accomplished with the use of well crafted unitary transformations before any RWA is performed \cite{Cessa, Jonathan,interfer}. 
Given its use in the next section, let us briefly review the approach presented in Ref. \cite{Jonathan}.
We will expand on the more general approach proposed by Ref. \cite{Cessa} in Sec. \ref{new}.

First of all, we consider the system in exact resonance, $\delta = 0$, in Eq. (\ref{eq:hamilt-0}). Using the coupling term in the Lamb-Dicke regime, Eq. (\ref{eq:hamilt-V-lb}), the Hamiltonian reads
\begin{equation} \label{eq:hamilt-jon0}
    H = \nu \adag a + \frac{\Omega}{2} \sigx + \frac{i \eta \Omega}{2} (\sigp - \sigm) (a + \adag).
\end{equation}
By applying the unitary transformation
\begin{equation} \label{eq:unit-r}
    R = \frac{1}{\sqrt{2}}\left(\begin{matrix}
        1 & 1 \\
        -1 & 1
    \end{matrix}\right)
\end{equation}
to Eq. (\ref{eq:hamilt-jon0}), one arrives at
\begin{equation}
    H' \equiv R H \dagg R = \nu \adag a + \frac{\Omega}{2} \sigz + \frac{i \eta \Omega}{2} (\sigp - \sigm) (a + \adag).
\end{equation}
Now, by choosing $\Omega = \nu$, one can perform the RWA and find another JCM-type effective Hamiltonian that reads
\begin{equation} \label{eq:hamilt-jon1}
    H'_{\rm{JCM}}\approx \nu \adag a + \frac{\Omega}{2} \sigz + \frac{i \eta \Omega}{2} (\sigp a - \sigm \adag).
\end{equation}

With effect, notice that this RWA requires $\eta \Omega \ll \nu$ to be valid, instead of $\Omega \ll \nu$, as in the derivation of Eq. (\ref{eq:hamilt-rsb}).
Since we are within the Lamb-Dicke regime with $\eta\sim 10^{-1}$, we can now work with $\Omega$ at least one order of magnitude higher than what was allowed before. One could argue that there is no gain in the speed of the time evolution given that the coupling constant in Eq. (\ref{eq:hamilt-jon1}) is $\eta\Omega$ and not $\Omega$. This view is misleading because, ultimately, the interaction Hamiltonian in Eq. (\ref{eq:hamilt-V}) is the one microscopically deduced and therefore communicates directly with experimental results.
It is also worth noticing that the Hamiltonian in Eq. (\ref{eq:hamilt-jon1}) lies in a transformed space and it is seen as just a useful tool to perform analytical calculations which must be transformed back to the original space in order to make physical predictions. 

\section{Protocol for determination of the motional Wigner function revisited}\label{wigner}

In this section we will give a concise review of the method of characterization of motional state proposed in Ref. \cite{wine}.
The method was conceived with the use of sideband Hamiltonians, such as the one in Eq. (\ref{eq:hamilt-rsb}).
We will also present an extension to this method, which, in turn, will allow the use stronger coupling constants $\Omega$.

The scenario considered is as follows.
Suppose that, at the end of a certain trapped ion protocol, one arrives at the state
\begin{equation} \label{eq:wig-init}
   \ket{\psi_0} = \ket{g} \otimes \ket{\phi},
\end{equation}
where the motional state $\ket{\phi}$ is unknown, and $\ket{g}$ is the eigenstate of $\sigma_z$ with eigenvalue $-1$, the electronic ground state.
The protocol proposed in Ref. \cite{wine} allow us to determine the motional state $\ket{\phi}$.

The first step is the application of the displacement operator $\Ddag(\alpha)$ to the initial state in Eq. (\ref{eq:wig-init}), that is
\begin{equation}
    |\td \psi_0 \rangle\equiv \ket{g} \otimes \Ddag(\alpha) \ket{\phi}.
\end{equation}
By allowing this state to evolve in time with the red sideband Hamiltonian in Eq. (\ref{eq:hamilt-rsb}), one finds that the ground state probability is given by
\begin{equation} \label{eq:prob-g}
    P_g(t) = \frac{1}{2}\left( 1 + \sum_{k=0}^\infty Q_k(\alpha) \cos(\eta \Omega \sqrt{k} t) \right),
\end{equation}
where $Q_k(\alpha) \equiv \abs{\langle k \vert \Ddag(\alpha) \vert \psi_0 \rangle}^2$.
After experimental determination of  $P_g(t)$, one can fit the curve and the coefficients $Q_k(\alpha)$ can be obtained.

With the coefficients $Q_k(\alpha)$ at hand, the relation \cite{cahilglauber}
\begin{equation} \label{eq:wigner-qk}
    W(\alpha) = \frac{2}{\pi} \sum_{k=0}^\infty (-1)^k Q_k(\alpha),
\end{equation}
allows then for the determination of the Wigner function of $\ket{\phi}$ at a given point $\alpha$.
Finally, by repeating this procedure throughout the phase space, the Wigner function of the motional state can be determined.
More details on the protocol can be found in Ref. \cite{wine}.
They also present a way of directly determining the density operator of the motional state by using these same quantities, $Q_k(\alpha)$, obtained for several phase space points, $\alpha$.

\subsection{Adapting the protocol for stronger lasers} \label{sec:wigner-fast}

In order to adapt this method for faster regimes,  one must go beyond the usual sideband Hamiltonians [Eq. (\ref{eq:hamilt-rsb})], whose validity is conditioned to the use of weak lasers, \emph{i.e.},  $\Omega\ll\nu$.
To that effect, we propose the use of the Hamiltonian in Eq. (\ref{eq:hamilt-jon1}), which works in the intermediate intensity regime, $\Omega \sim \nu$.
This regime will be carefully investigated with numerical methods in Sec. \ref{new}.
As a matter of fact, in that section we will present a new effective Hamiltonian, which is valid for a broader range of physical parameters when compared to Eq. (\ref{eq:hamilt-jon1}).
In the context of protocols for Wigner function reconstruction, however, the use of such a Hamiltonian would bring unnecessary complications concerning the observables to be measured in place of the much simpler $P_g(t)$.
The reason is that the overall unitary transformation we will use to obtain such a Hamiltonian is much more complicated than that in Eq. (\ref{eq:unit-r}), causing entanglement between projectors acting on motion and internal degrees of freedom.
Given all that, for the purpose of the Wigner function protocol, we will focus on the Hamiltonian in Eq. (\ref{eq:hamilt-jon1}).

The protocol as presented in the last section asks for a JCM-like Hamiltonian, which is exactly what we have with Eq. (\ref{eq:hamilt-jon1}).
Nevertheless, it is important to remember that this effective Hamiltonian has this form in a frame rotated by the unitary transformation $R$ [Eq. (\ref{eq:unit-r})].
Since the actual experiment takes place in an unrotated frame, we must be consistent with which basis we are considering.
For clarity, let us name the frame rotated by $R$ as ``JCM frame'' and the usual, unrotated frame, as ``Lab frame''.

At the start of the protocol, we need the state to have the form of Eq. (\ref{eq:wig-init}) in the JCM frame.
This can be accomplished, in the Lab frame, by the use of a rotated initial state
\begin{equation}\label{eq:wig-init-r}
    \ket{\psi_0}_\tx{LAB} = \dagg R \ket{g} \otimes \ket{\phi} = - \ket{-} \otimes \ket{\phi},
\end{equation}
where $\ket{-} = (\ket{e}-\ket{g})/\sqrt{2}$ is the eigenstate of $\sigma_x$ corresponding to the eigenvalue $-1$.
Thus, in the JCM frame, the initial state will read
\begin{equation}
    \ket{\psi_0}_\tx{JCM} = R \ket{\psi_0}_\tx{LAB} = \ket{g} \otimes \ket{\phi},
\end{equation}
as needed.

The next step is to apply the displacement operator $\Ddag(\alpha)$ in the JCM frame.
This step remains unchanged in the Lab frame, since
\begin{equation} \label{eq:wig-disp-r}
    \dagg R \Ddag(\alpha) R = \Ddag(\alpha).
\end{equation}
Now the system is let to evolve according to the effective Hamiltonian in Eq. (\ref{eq:hamilt-jon1}), and $\Pi_g = \ketbra{g}{g}$ is measured in the JCM frame.
This amounts to measuring
\begin{equation} \label{eq:wig-proj-r}
    \dagg R \Pi_g R = \ketbra{-}{-}
\end{equation}
in the Lab frame, which is a simple measurement acting only on the electronic degrees of freedom.
It corresponds to the probability of finding the ion in the electronic state $\ket{-}$ at a given instant, $t$, which we denote as $P_{-}(t)$.
This probability can be determined from the known $P_g(t)$ [Eq. (\ref{eq:prob-g})] for a initial state of the form Eq. (\ref{eq:wig-init}) evolving according to a JCM-like Hamiltonian, since

\begin{align}
    P_-^\tx{LAB}(t) &= |\braket{-}{\psi(t)}_\tx{LAB}|^2 = |\braket{g}{\psi(t)}_\tx{JCM}|^2 = P_g^\tx{JCM}(t) \nonumber \\
    &= \frac{1}{2}\left( 1 + \sum_{k=0}^\infty Q_k(\alpha) \cos(\eta \Omega \sqrt{k} t) \right)
\end{align}

Finally, one proceeds to fit the curve $P_{-}(t)$ to determine the quantities $Q_k(\alpha)$ in order to calculate $W(\alpha)$ [Eq. (\ref{eq:wigner-qk})], just like in the original protocol.
The  same steps are then repeated for each chosen phase space point, $\alpha$.
A summary of the objects, both in the Lab and JCM frame, important to this generalized protocol is presented in Table \ref{tab:table}.

Our guide in adapting this protocol for stronger lasers was to keep the operations as simple as possible.
Had we considered the more general unitary transformation, which will be presented in a later section, we would not be able to use the simple objects in Eqs. (\ref{eq:wig-init-r}), (\ref{eq:wig-disp-r}) and (\ref{eq:wig-proj-r}).
Since this general transformation tends to entangle electronic and motional degrees of freedom, the objects used in the Lab frame would become considerably more complicated.

\begin{widetext}
    \begin{minipage}{\textwidth}
        \begin{table}[H]
            \centering
            \normalsize
            \begin{tabular}{|c|c|c|}
                \hline
                & \tb{JCM frame} & \tb{Lab frame} \\
                \hline
                \tb{Initial state} & $\ket{\psi_0} = \ket{g} \otimes \ket{\phi}$ & $\ket{\psi_0} = - \ket{-} \otimes \ket{\phi}$ \\
                \hline
                \tb{Disp. operator} & $\Ddag(\alpha)$ & $\Ddag(\alpha)$ \\
                \hline
                \tb{Measurement} & $P_g(t) = \frac{1}{2}\left( 1 + \sum_{k=0}^\infty Q_k(\alpha) \cos(\eta \Omega \sqrt{k} t) \right)$ & $P_-(t) = \frac{1}{2}\left( 1 + \sum_{k=0}^\infty Q_k(\alpha) \cos(\eta \Omega \sqrt{k} t) \right)$ \\
                \hline
            \end{tabular}
            \caption{Summary of the objects needed in the Wigner function determination protocol in both frames.}
            \label{tab:table}
        \end{table}
    \end{minipage}
\end{widetext}

\subsection{Inclusion of decoherence effects}

Let us suppose that a target motional state $\ket{\phi}$ is produced in a particular trapped ion experiment, and one wants to characterize it by applying either the original protocol \cite{wine} or the one proposed here, which goes beyond the sideband Hamiltonians.
Assuming that there is no limitations to the number of phase space points used to reconstruct the Wigner function, both protocols would work perfectly well in the idealized scenario where no decoherence takes place.
The next step we take, then, is to phenomenologically include some decoherence effect.

It is not our goal to realistically model the decoherence and noise present in trapped ion setups.
This is an important problem in itself which has received increasing attention over the last years \cite{decoh1,decoh2,decoh3,decoh4,decoh5,decoh6}.
Instead, what we would like here is to come up with a toy model able to highlight the need to design fast protocols for quantum technologies and to motivate similar studies with more realistic models.

In this spirit, we consider that the free evolution step in the protocol introduced earlier this section will be subjected to a dephasing channel.
Effectively, this will be accomplished by the inclusion of a noise term of the form
\begin{equation} \label{eq:deph-lab}
    \D_\tx{deph}(\rho) = \frac{\gamma}{2} \bigl( \sigz \rho \sigz - \rho \bigr)
\end{equation}
in the master equation, where $\gamma$ is the dephasing rate.
This term is written in the Lab frame.
In order to describe how the evolution in the JCM frame will be affected, we again use the transformation $R$ [Eq. (\ref{eq:unit-r})], arriving at
\begin{equation} \label{eq:deph-jcm}
    \D_\tx{deph}^\tx{JCM}(\rho_\tx{JCM}) = \frac{\gamma}{2} \bigl( \sigx \rho_\tx{JCM} \sigx - \rho_\tx{JCM} \bigr).
\end{equation}
We proceed to numerically demonstrate the need for the faster version of the protocol in scenarios of slightly larger values of the dephasing rate, $\gamma$.

\subsection{Numerical support} \label{sec:numerics}

In order to underpin the importance of exploring faster regimes in the trapped ion setup, we now present some numerical results.
All the numerics presented were done using the QuantumOptics.jl framework \cite{qojulia} for the Julia programming language \cite{julialang}.

For comparison, in each case described later, we ran two simulations. One following the protocol in the low intensity regime \cite{wine}. And the other following the adapted protocol for the intermediate intensity regime, as described earlier in this section.
In both cases the steps were i) initialization of a chosen state $\ket{\psi_0}$ [Eq. (\ref{eq:wig-init})] to be determined; ii) application of the displacement operator $\Ddag(\alpha)$; iii) evolution of the probability $P_g(t)$ according to the respective master equation, depending on the regime in case; iv) fitting of the data calculated numerically to the analytical expected result [Eq. (\ref{eq:prob-g})] and determination of the coefficients $Q_k(\alpha)$; v) finally, with this information we can calculate the Wigner function at point $\alpha$ [Eq. (\ref{eq:wigner-qk})].
These steps are then repeated for each chosen point $\alpha$.
Notice that the fitting of the numerical data will introduce errors in the calculation of $Q_k(\alpha)$ beyond those naturally expected for numerical calculations.
This is due to the fact that the expression used to fit the data [Eq. (\ref{eq:prob-g})] was derived considering that the system is evolving coherently.

For convenience, in each case we performed the calculations in the JCM frame.
In the low intensity regime, the system evolved according to the red sideband Hamiltonian in Eq. (\ref{eq:hamilt-rsb}), with parameters set to $\delta = \nu$ and $\Omega = 0.05 \nu$.
For a typical ion trap with secular frequency in the order of $\nu = 1 \tx{MHz}$ \cite{review}, this would mean a Rabi frequency of about $\Omega = 50 \tx{KHz}$.
For the intermediate intensity regime, the system evolved according to the effective Hamiltonian in Eq. (\ref{eq:hamilt-jon1}), with $\delta = 0$ and $\Omega = \nu$.
From the experimental side, this regime of $\Omega$ comparable to $\nu$ can be achieved using, for example, a trapped Yb ion coupled to a fiber Fabry-Perot cavity \cite{fabry-perot}.
In both cases the Lamb-Dicke parameter was set to $\eta = 0.05$ and the cut-off of the number basis for the motional subspace was set to $N=50$.
Also in both cases, the time evolution step was performed for a total time of $\Omega t = 800$ for each point $\alpha$ considered.
Notice that, since the magnitude of $\Omega$ is different in each of the regimes considered, the times in the more comparable units $\nu t$ will be much larger for the low intensity regime.

As stated before, we introduced a dephasing term in the evolution step in order to model some decoherence effect interfering with the protocol.
Besides the parameter values, this is the only other point where the simulations diverge.
Since the dephasing term Eq. (\ref{eq:deph-lab}) is written in the Lab frame, it will only act with this form in the low intensity case, since the sideband Hamiltonian Eq. (\ref{eq:hamilt-rsb}) is already written in this frame.
In the intermediate intensity case, since the JCM and Lab frames are not the same, we have to use the transformed version of the noise term in Eq. (\ref{eq:deph-jcm}).
We performed the simulations for increasing values of the dephasing rate $\gamma$ in the interval $0.0004 \nu \leq \gamma \leq 0.05 \nu$.
Again, considering a secular frequency of order $\nu \sim 1 \tx{MHz}$, the dephasing rates would be in the interval $0.4 \tx{KHz} \lesssim \gamma \lesssim 50 \tx{KHz}$.

As initial motional states we chose some recognizable cases.
Namely, number, coherent, and cat states.
In every case the results were qualitatively similar in that the faster protocol always gave better results when compared to the slower one, with the difference being more prominent the larger the dephasing rate $\gamma$.

\begin{figure}
    \centering
    \includegraphics[width=0.48\textwidth]{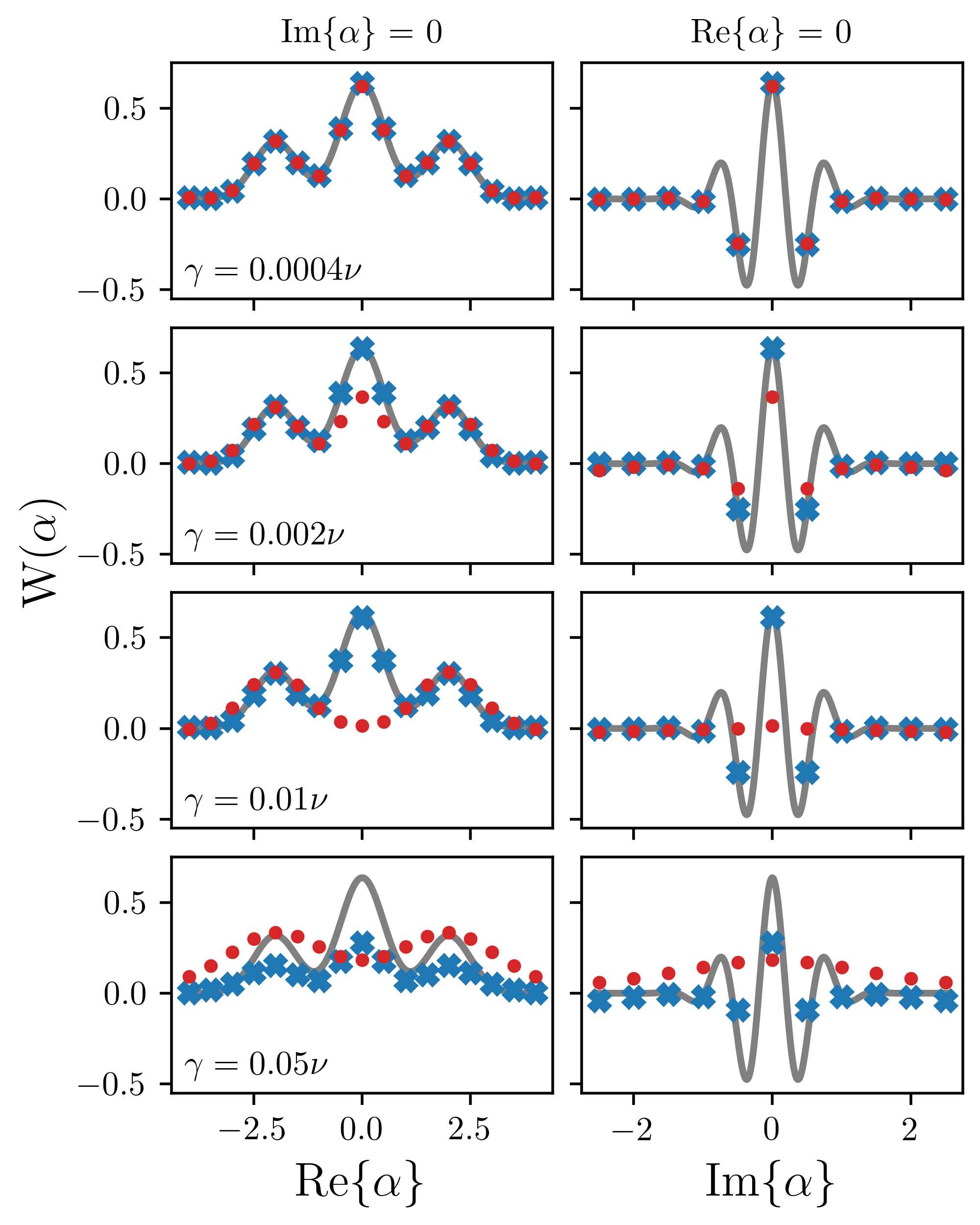}
    \caption{(Color online) Numerical simulation of the Wigner function determination protocol for a motional cat state [Eq. (\ref{cats})] in the presence of dephasing. In each panel we present the analytical Wigner function (gray solid line), the results using the slow regime protocol (red dots), and the results using our proposed faster protocol (blue Xs). Each column gives a specific slice of the Wigner function. And with each row the value of the dephasing rate $\gamma$ is increased.}
    \label{fig:wig-deph-cat}
\end{figure}

To illustrate this, in Fig. \ref{fig:wig-deph-cat} we present the results for the simulation using a motional cat state
\begin{equation}\label{cats}
\ket{\tx{cat}}=\mathcal{N} (\ket{\alpha = 2} + \ket{\alpha = -2}),
\end{equation} 
where $\mathcal{N}$ is a normalization factor.
Since the full Wigner function would require a 3D plot, we opted for a slicing of the phase space.
Namely, we calculated the Wigner function in the planes with $\tx{Im}\{\alpha\}=0$ (left column of Fig. \ref{fig:wig-deph-cat}) and $\tx{Re}\{\alpha\}=0$ (right column).
In each row of the figure is a run of the protocol with different dephasing rates $\gamma$.
The analytical form of the Wigner function for the cat state is presented as a gray solid line for comparison.
The red dots are the points returned by the protocol in the slower regime and the blue Xs, the points returned by the faster version.

As the figure shows, the faster protocol visually outperformed the slower one in each run, even if only slightly in the case with least dephasing acting ($\gamma = 0.0004 \nu$).
As stated before, the difference becomes more visible with increasing $\gamma$.
Already in the case $\gamma = 0.01\nu$, the slower protocol returned a completely ruined figure, with only the two bumps visible in the left slice, but without the interference pattern in the center or the negative values expected in the right slice.
The protocol adapted for the faster regime returned acceptable results for dephasing rates up to $\gamma = 0.01\nu$.
And even in the extreme case $\gamma = 0.05\nu$, the result was able to keep a general shape and even some negativity in the right slice.
Even with this simple example, we hope to convince the reader of some of the advantages of exploring faster regimes in the trapped ion setup.

\section{Beyond sideband Hamiltonians in the case of non-resonant interaction}\label{new}

As a second result of this work, we present a new effective Hamiltonian able to give accurate results in the intermediate intensity regime, $\Omega\sim\nu$, even in the case of non-resonant laser interactions, $\delta\neq 0$.
Let us go back to the complete Hamiltonian, defined in Eq. (\ref{eq:hamilt-complete}), upon which the following unitary transformation \cite{Cessa,cessa-review}
\begin{equation} \label{eq:unit-t}
    \begin{split}
        T(\beta) &= \frac{1}{2 \sqrt{2}} \biggl( (I + \sigz) \Ddag(\beta) + (I - \sigz) \D(\beta) + \\
        &\quad + 2 \sigp \D(\beta) - 2 \sigm \Ddag(\beta) \biggr) \\
        &=\frac{1}{\sqrt{2}}\left(\begin{matrix}
            \Ddag(\beta) & \D(\beta)\\
            -\Ddag(\beta) & \D(\beta)
        \end{matrix}\right),
    \end{split}
\end{equation}
is applied with $\beta = i \eta / 2$.
Notice, as stated before, how this generalized transformation entangles electronic and motional degrees of freedom, as opposed to the simpler Eq. (\ref{eq:unit-r}).
The transformed Hamiltonian then reads
\begin{equation} \label{eq:hamilt-mc1}
    \begin{split}
        H' &\equiv T H \dagg{T} \\
        &= \nu \dagg a a + \frac{\Omega}{2} \sigz  - \frac \delta 2 \sigx + \frac{i \eta \nu}{2} \sigx (a - \dagg a).
    \end{split}
\end{equation}
The next step according to Ref. \cite{Cessa} is the application of a RWA with the resonance condition $\Omega = \nu$, arriving at
\begin{equation} \label{eq:hamilt-mc}
    H'_\tx{MC} \approx \nu \adag a + \frac{\Omega}{2} \sigz + \frac{i \eta \nu}{2} (\sigp a - \sigm \adag),
\end{equation}
which is a JCM-like effective Hamiltonian akin to the one in Eq. (\ref{eq:hamilt-jon1}). What is to be remarked here is that such a RWA is unlikely to give accurate results when $\delta\neq 0$. Let us consider, for example, the usual choice $\delta=\nu$. A close look at Eq. (\ref{eq:hamilt-mc1}) reveals that the  $\sigma_x$ term would appear with a coupling constant with magnitude $\nu/2$ which is stronger than $\eta \nu/2$. The result is that the $\sigma_x$ term can not be neglected for $\delta \neq 0$ as done in Ref. \cite{Cessa}. In fact, as we will see later, their effective Hamiltonian starts losing accuracy as $\delta$ increases.

Having that in mind, we propose here a further step prior to the RWA, which will allow one to obtain accurate results for a slightly broader range of detunings $\delta$. This step is the diagonalization of the two-level free term in Eq. (\ref{eq:hamilt-mc1}), \emph{i.e.},
\begin{equation}
    U \left( \frac{\Omega}{2} \sigz - \frac{\delta}{2} \sigx \right) \dagg U = \frac{\td \omega}{2} \sigz,
\end{equation}
by means of the unitary transformation
\begin{equation}
    U = \frac{1}{\sqrt{2 \td \omega}}\left(\begin{matrix}
        \sqrt{\td \omega + \Omega} & - \sqrt{\td \omega - \Omega} \\
        \sqrt{\td \omega - \Omega} & \sqrt{\td \omega + \Omega}
    \end{matrix}\right)
\end{equation}
where
\begin{equation} \label{eq:omega-td-def}
    \td \omega \equiv \sqrt{\Omega^2 + \delta^2}.
\end{equation}
Applying $U$ to the full Hamiltonian [Eq. (\ref{eq:hamilt-mc1})], one finds
\begin{equation}
    \begin{split} \label{eq:hamilt-ts}
        H'' &= U T H \dagg T \dagg U \\
        &= \nu \adag a + \frac{\td \omega}{2} \sigz - \frac{i \delta \eta \nu}{2 \td \omega} \, \sigz (a - \adag) + \\
        &\quad + \frac{i \Omega \eta \nu}{2 \td \omega} \, \sigx (a - \adag),
    \end{split}
\end{equation}
and, only now, we make a RWA, arriving at the JCM-like effective Hamiltonian
\begin{equation} \label{eq:hamilt-ts-rwa}
    H'' = \nu \adag a + \frac{\td \omega}{2} \sigz + \frac{i \Omega \eta \nu}{2 \td \omega} (\sigp a - \sigm \adag).
\end{equation}
with resonance condition given by $\td \omega  = \nu$.

One interesting aspect of this resonance condition is that it is valid for a whole range of values of the parameters $\delta$ and $\Omega$, which represents a significant improvement over previous approaches  \cite{Cessa,Jonathan,interfer}.
Following from the definition of $\td \omega$ [Eq. (\ref{eq:omega-td-def})], and from the new resonance condition, $\td \omega  = \nu$, one finds
\begin{equation}\label{casesr}
\Omega=\sqrt{\nu^2-\delta^2},
\end{equation}
with
$0 \leq \abs{\delta} < \nu$.
Therefore, our effective Hamiltonian in Eq. (\ref{eq:hamilt-ts-rwa}) functions as a bridge between the well-known sideband Hamiltonians valid at the low ($\Omega \ll \nu$ and $\delta \sim \nu$) and the intermediate  intensity regimes ($\Omega \sim \nu$ and $\delta \sim 0$). This interpolation was prohibited in the previous approaches given that, contrary to Eq. (\ref{casesr}), they had to fix $\Omega$ in a resonance condition which did not involved $\delta$.

Now, one final comment concerning the validity of Hamiltonian in Eq. (\ref{eq:hamilt-ts-rwa}). When passing from Eq. (\ref{eq:hamilt-ts}) to Eq. (\ref{eq:hamilt-ts-rwa}), we neglected the terms
\begin{equation}\label{e1}
    \frac{i \delta \eta \nu}{2 \wt} \sigz(a - \adag)
\end{equation}
and
\begin{equation}\label{e2}
    \frac{i \Omega \eta \nu}{2 \wt} (\sigm a - \sigp \adag).
\end{equation}
As said before, the validity of the RWA relies on the magnitude of the coupling constants relative to $\nu$.
Given that the resonance condition requires $\wt = \nu$, the important quantities to analyze in Eqs. (\ref{e1}) and (\ref{e2}) are then $\delta \eta$ and $\Omega \eta$, respectively.
Since our main goal is to provide analytical progress in regimes where $\Omega \sim \nu$, the need for $\Omega\eta\ll\nu$ will impose $\eta\ll 1$. This is not a problem, since the experiments usually operate in the Lamb-Dicke regime \cite{review}. Besides, the condition $\eta \ll 1$ also contributes in satisfying $\delta \eta \ll \nu$.
However, one must pay attention to the case of small $\Omega$, \emph{i.e.}, $\Omega\ll\nu$. According to Eq. (\ref{casesr}), the detuning $\delta$ must approach $\nu$ in that case. This would tend to increase the importance of the term in Eq. (\ref{e1}). Again, this is not a problem, since the sideband Hamiltonians work well in this low intensity regime \cite{review}.

\subsection{Numerical support} 

Now we present some numerical simulations to check the validity of the effective Hamiltonian in Eq. (\ref{eq:hamilt-ts-rwa}). We will be fixing a initial state, and then let it evolve according to our result, \emph{i.e.}, the Hamiltonian in Eq. (\ref{eq:hamilt-ts-rwa}), the red sideband Hamiltonian in Eq. (\ref{eq:hamilt-rsb}) and the effective Hamiltonian in Eq. (\ref{eq:hamilt-mc}). We do not include simulations with the Hamiltonian in Eq. (\ref{eq:hamilt-jon1}) given that it is just the one in Eq. (\ref{eq:hamilt-mc}) in the Lamb-Dicke regime and $\delta=0$.
At each time step, we evaluate the fidelity
\begin{equation}
    \mathcal{F}(t) = \abs{\braket{\td \psi(t)}{\psi(t)}}^2,
\end{equation}
where $\ket{\psi(t)}$ is the state numerically evolved with the complete Hamiltonian in Eq. (\ref{eq:hamilt-complete}) and $\ket{\td \psi(t)}$ is the state obtained with one of the effective Hamiltonians in Eqs. (\ref{eq:hamilt-rsb}), (\ref{eq:hamilt-mc}), and (\ref{eq:hamilt-ts-rwa}).

It is important to mention that, for the total evolution times considered ($\nu t$ well into the hundreds or even thousands, as will be seen shortly), the curve of the fidelity $\mathcal{F}$ presents extremely fast, small oscillations around a well defined average curve.
Presenting all this information here would be fruitless, since the oscillations blur the lines in the graphs.
So, in order to enhance the intelligibility, we averaged over these oscillations and will present this quantity.
Rigorously, what was done was the averaging of the fidelity values $\mathcal{F}(t)$ over time intervals of $\Omega T = 2 \pi$.
This can be understood as a coarse-graining of the time parameter.

Again, we performed the simulations for an assortment of initial states, always leading to qualitatively similar results.
For the results presented here, we chose a toy state that includes entanglement and different motional states
\begin{equation}
    \ket{\psi_0} = \mathcal{N} \bigl(\ket{e} \otimes \ket{\alpha = 2} + \ket{g} \otimes \ket{n = 4}\bigr),
\end{equation}
where $\mathcal{N}$ is a normalization factor, $\ket{n}$ denotes a motional number state, and $\ket{\alpha}$ denotes a motional coherent state.

\begin{figure}
    \centering
    \includegraphics[width=0.43\textwidth]{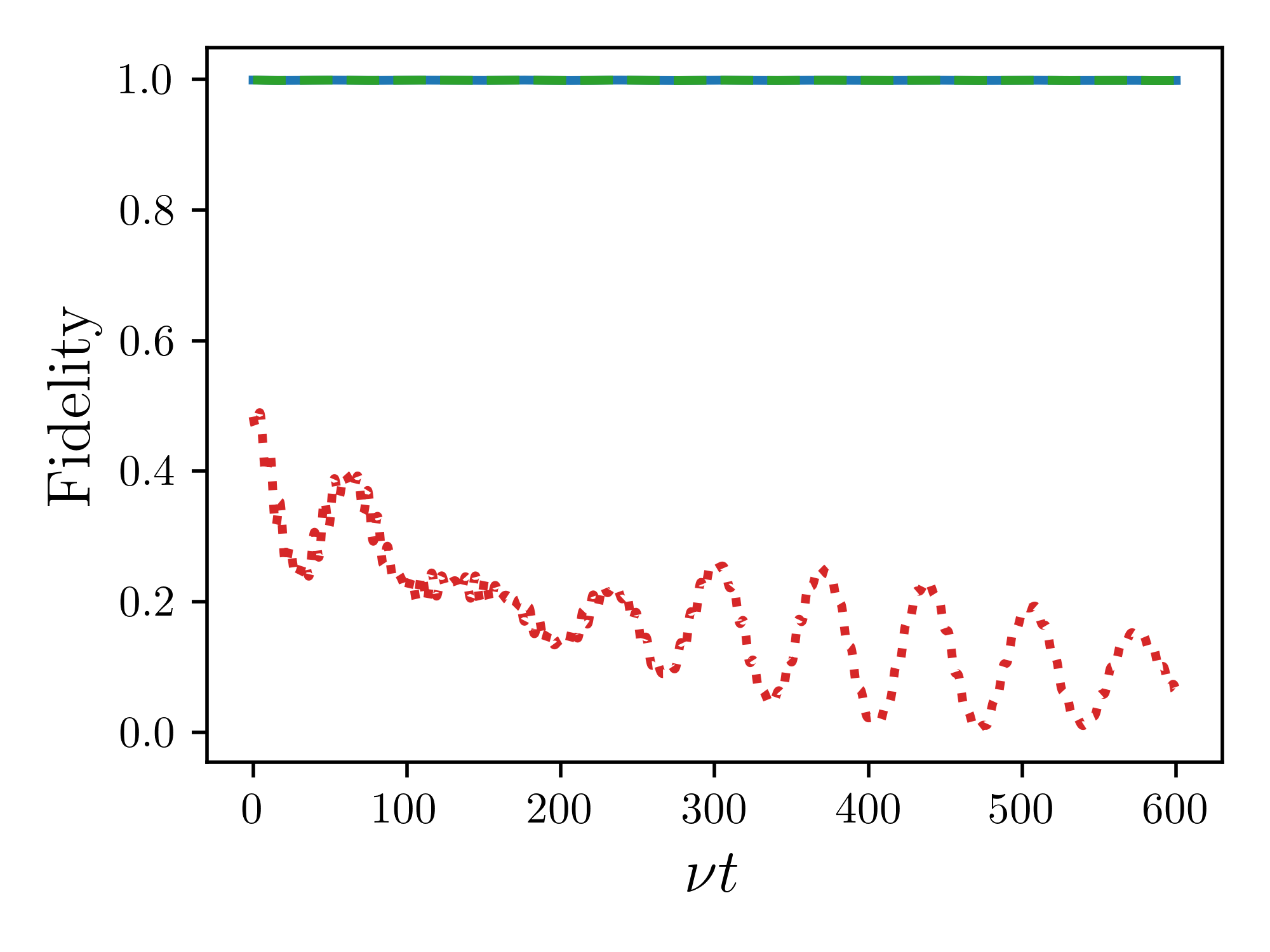}
    \caption{(Color online) Numerical evolution of the fidelity between a state evolved according to effective Hamiltonians in Eqs. (\ref{eq:hamilt-rsb}) (red dotted line), (\ref{eq:hamilt-mc}) (green dashed line), and (\ref{eq:hamilt-ts-rwa}) (blue solid line), and one evolved according to the complete Hamiltonian in Eq. (\ref{eq:hamilt-complete}). Results for the intermediate intensity regime in the resonant case, with parameters $\Omega = 1.0 \nu$ and $\delta = 0$.}
    \label{fig:fid-fast1}
\end{figure}

Starting at the faster regime, in Fig. \ref{fig:fid-fast1} we present the results using parameters $\Omega = 1.0 \nu$ and $\delta = 0$. As expected, the effective Hamiltonian in Eqs. (\ref{eq:hamilt-mc}) (green dashed line) and (\ref{eq:hamilt-ts-rwa}) (blue solid line) produced identical results, since the former is a particular case of the latter for $\delta = 0$. Both of them kept the fidelities over $\mathcal{F} > 99\%$ during the whole time interval considered. Not surprisingly, the red sideband Hamiltonian (red dotted line) performs poorly, since we are far from its validity regime, which is the low intensity regime, $\Omega\ll\nu$.

\begin{figure}
    \centering
    \includegraphics[width=0.43\textwidth]{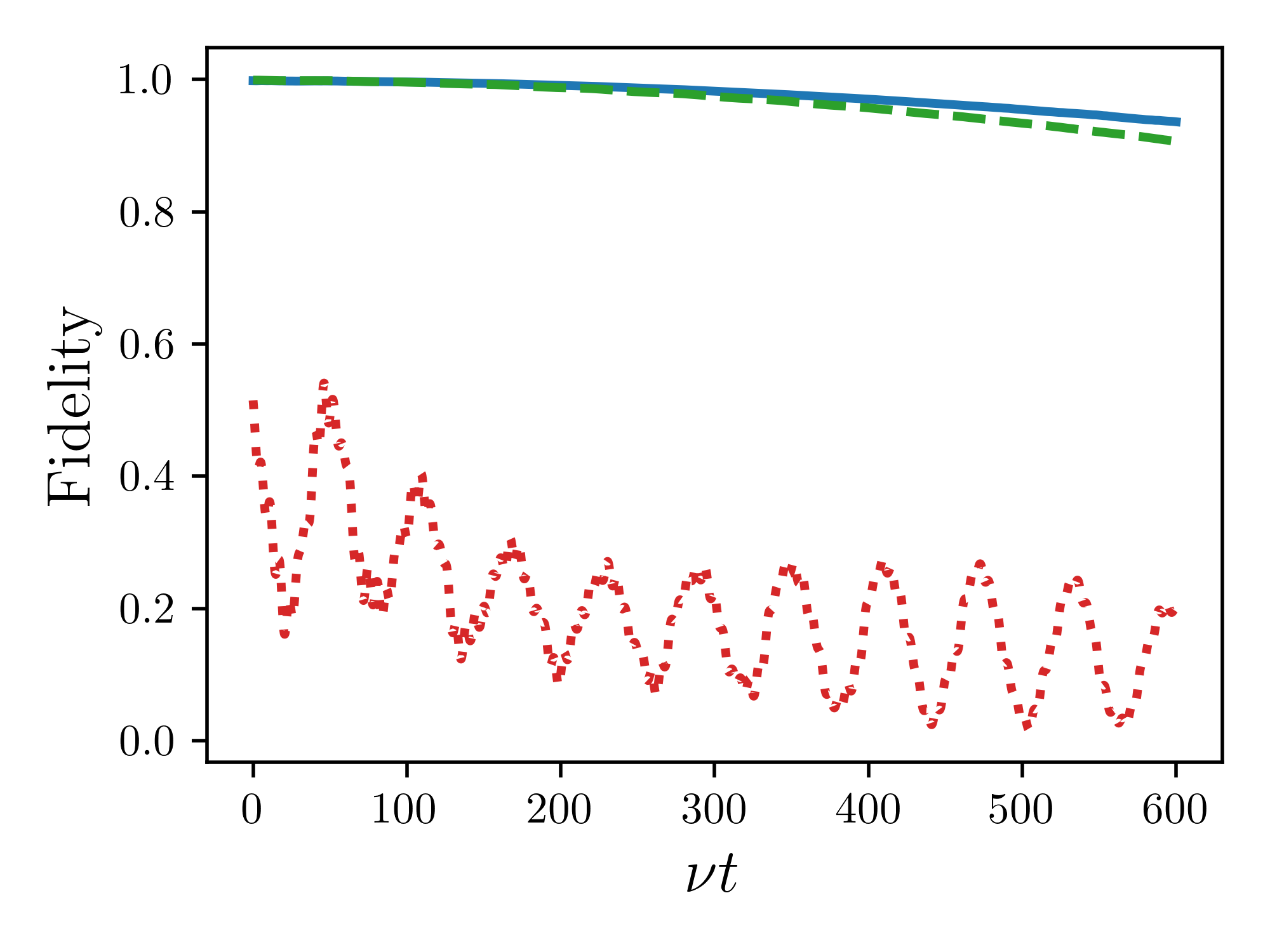}
    \caption{(Color online) Numerical evolution of the fidelity between a state evolved according to effective Hamiltonians in Eqs. (\ref{eq:hamilt-rsb}) (red dotted line), (\ref{eq:hamilt-mc}) (green dashed line), and (\ref{eq:hamilt-ts-rwa}) (blue solid line), and one evolved according to the complete Hamiltonian in Eq. (\ref{eq:hamilt-complete}). Results for the intermediate intensity regime in the non-resonant case, with parameters $\Omega = 1.0 \nu$ and $\delta = 0.3 \nu$.}
    \label{fig:fid-fast21}
\end{figure}

\begin{figure}
    \centering
    \includegraphics[width=0.43\textwidth]{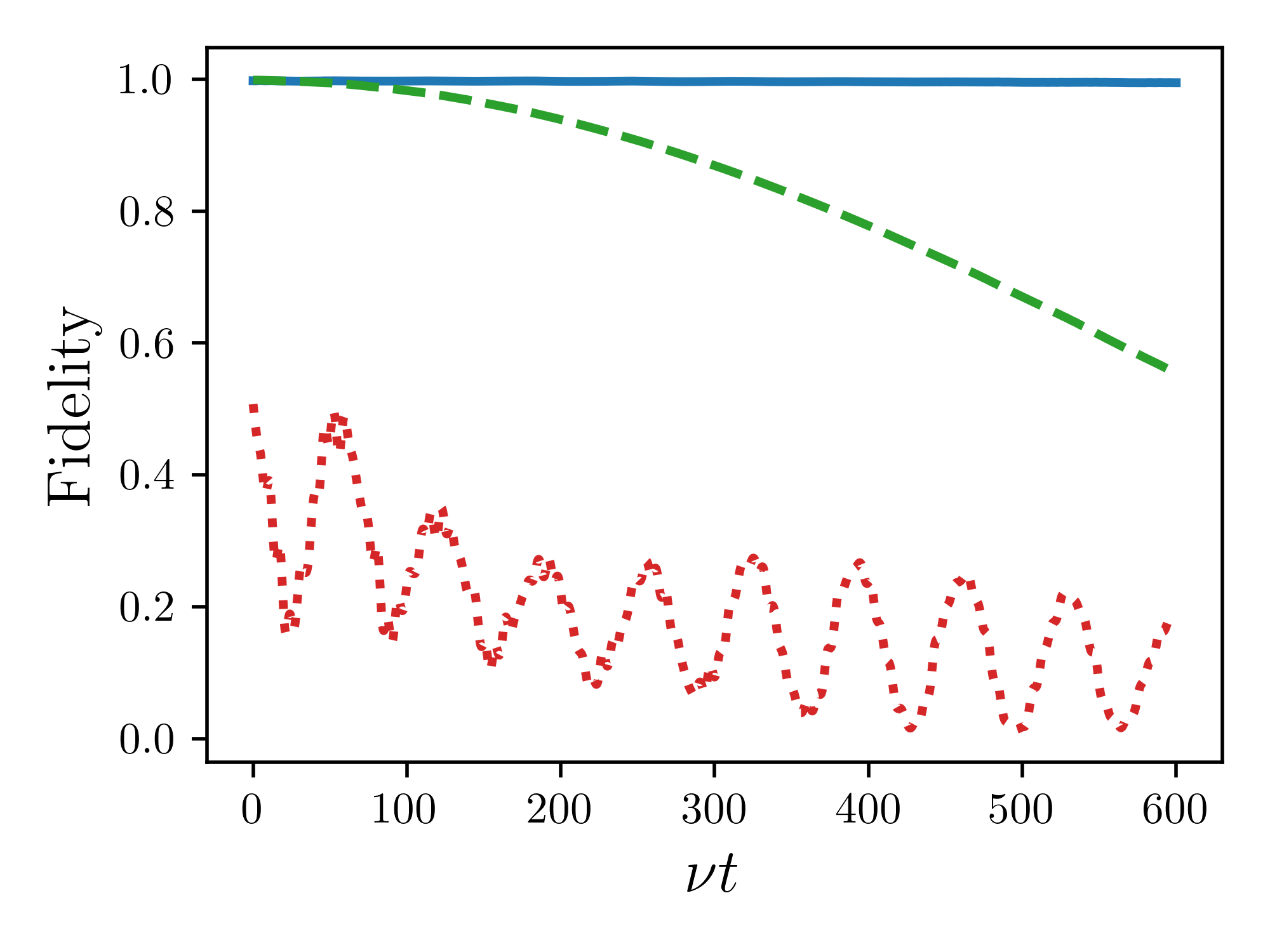}
    \caption{(Color online) Numerical evolution of the fidelity between a state evolved according to effective Hamiltonians in Eqs. (\ref{eq:hamilt-rsb}) (red dotted line), (\ref{eq:hamilt-mc}) (green dashed line), and (\ref{eq:hamilt-ts-rwa}) (blue solid line), and one evolved according to the complete Hamiltonian in Eq. (\ref{eq:hamilt-complete}). Results for the intermediate intensity regime in the non-resonant case, with parameters $\Omega = 0.95 \nu$ and $\delta = 0.3 \nu$.}
    \label{fig:fid-fast22}
\end{figure}

In Figs. \ref{fig:fid-fast21} and \ref{fig:fid-fast22} we illustrate a case of non-resonant interaction, \emph{i.e.}, $\delta \neq 0$. First, in Fig. \ref{fig:fid-fast21} we used $\Omega = 1.0 \nu$ and $\delta = 0.3 \nu$.
Neither Eq. (\ref{eq:hamilt-mc}) nor Eq. (\ref{eq:hamilt-ts-rwa})  performed particularly well, although the model proposed here performed slightly better, presenting $\mathcal{F} > 95\%$ for the total time considered.
It is worth noticing that our model was actually expected to not work well because $\delta = 0.3 \nu$ demands a retuning of the Rabi frequency $\Omega$ in order to satisfy the resonance condition $\wt = \nu$ or, equivalently, to satisfy Eq. (\ref{casesr}).
In Fig. \ref{fig:fid-fast22} we keep $\delta = 0.3 \nu$ and use $\Omega = 0.95 \nu$, which is much closer to the resonance condition in Eq. (\ref{casesr}).
This time around, we managed once more to keep $\mathcal{F} > 99\%$ over the total time duration.
At the same time, the result using Eq. (\ref{eq:hamilt-mc}) worsened, as $\Omega = 0.95 \nu$ is outside of its resonance condition $\Omega = \nu$.
It is then clear that Eq. (\ref{eq:hamilt-ts-rwa}) represents a step forward in the direction of having diagonalizable effective Hamiltonians covering new regimes of operation in trapped ions manipulated by lasers beams.

\begin{figure}
    \centering
    \includegraphics[width=0.43\textwidth]{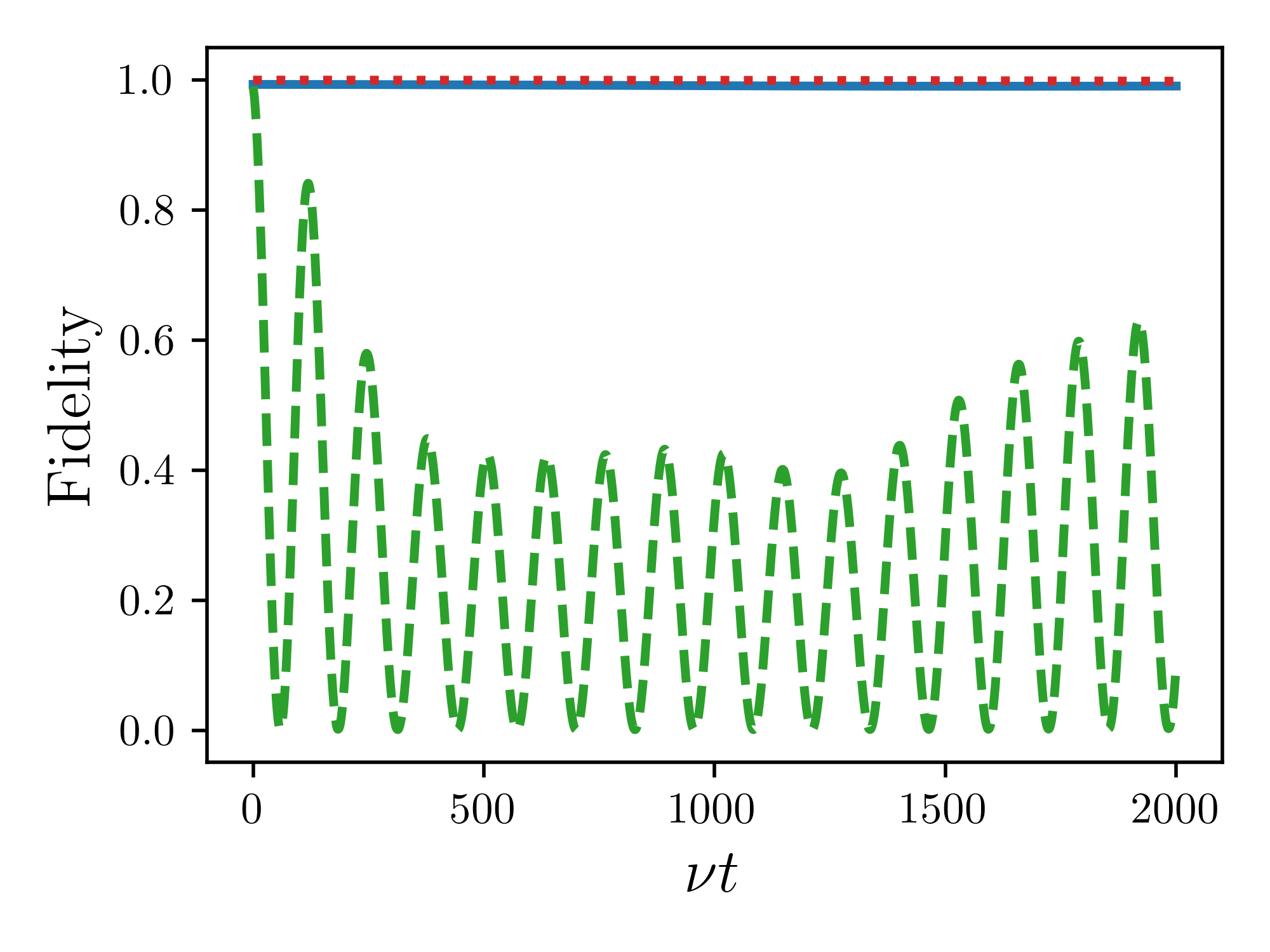}
    \caption{(Color online) Numerical evolution of the fidelity between a state evolved according to effective Hamiltonians in Eqs. (\ref{eq:hamilt-rsb}) (red dotted line), (\ref{eq:hamilt-mc}) (green dashed line), and (\ref{eq:hamilt-ts-rwa}) (blue solid line), and one evolved according to the complete Hamiltonian in Eq. (\ref{eq:hamilt-complete}). Results for the low intensity regime, with parameters $\Omega = 0.01 \nu$ and $\delta = 1.0 \nu$.}
    \label{fig:fid-slow1}
\end{figure}

\begin{figure}
    \includegraphics[width=0.43\textwidth]{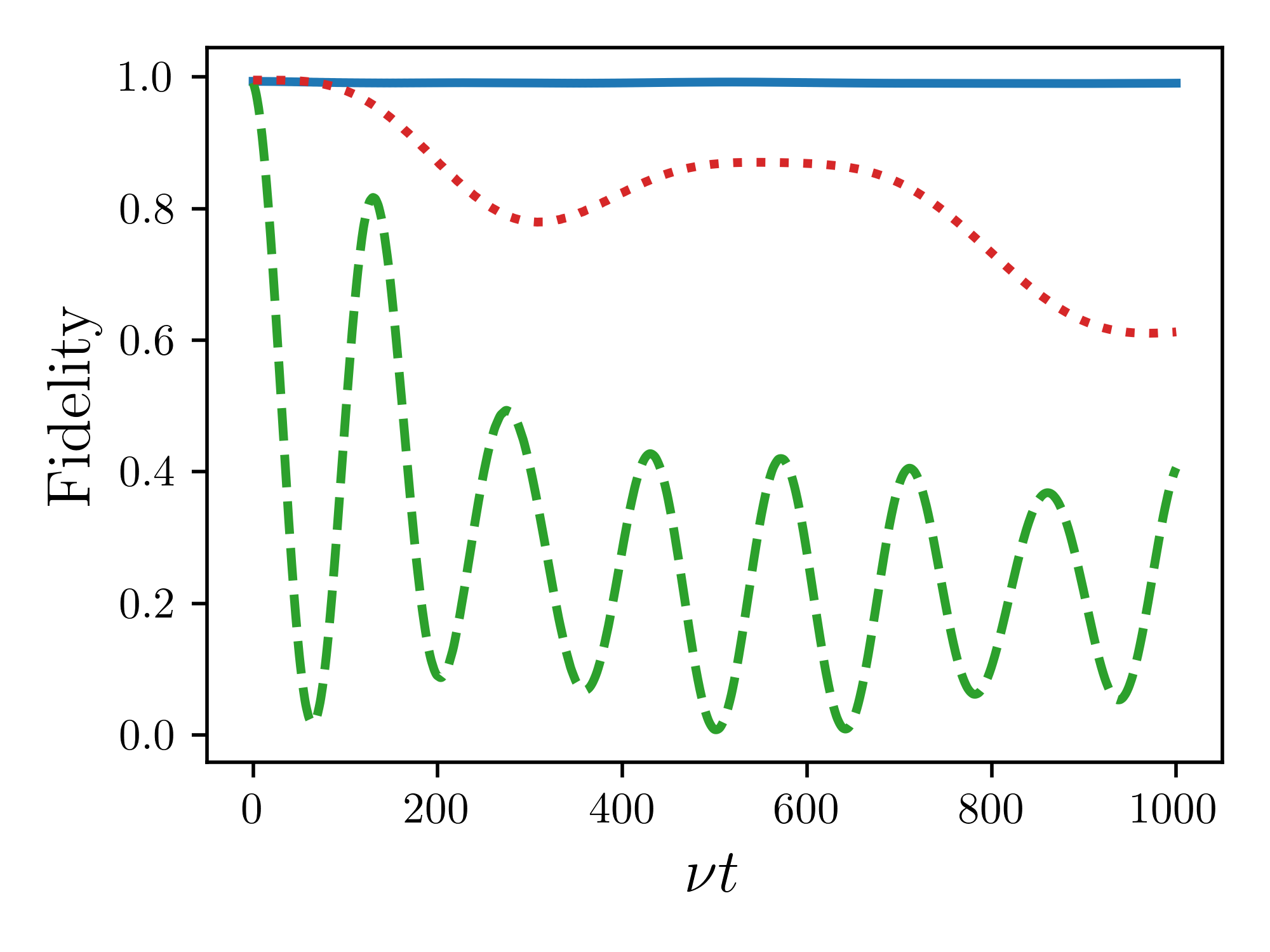}
    \caption{(Color online) Numerical evolution of the fidelity between a state evolved according to effective Hamiltonians in Eqs. (\ref{eq:hamilt-rsb}) (red dotted line), (\ref{eq:hamilt-mc}) (green dashed line), and (\ref{eq:hamilt-ts-rwa}) (blue solid line), and one evolved according to the complete Hamiltonian in Eq. (\ref{eq:hamilt-complete}). Results for the low intensity regime, with parameters $\Omega = 0.1 \nu$ and $\delta = 1.0 \nu$.}
    \label{fig:fid-slow2}
\end{figure}

Finally, we present in Figs. \ref{fig:fid-slow1} and \ref{fig:fid-slow2} examples in the low intensity regime ($\Omega \ll \nu$). Our goal is to check how well our model interpolates the weak and intermediate intensity regimes. For both panels we used $\delta = 1.0 \nu$.
In Fig. \ref{fig:fid-slow1}, we used $\Omega = 0.01 \nu$ and it is clear that the evolution according to the red sideband Hamiltonian in Eq. (\ref{eq:hamilt-rsb}) outperformed the others. The result using Eq. (\ref{eq:hamilt-mc}) failed as expected, since we are far from its validity region determined by the resonance $\Omega=\nu$.
Our model in Eq. (\ref{eq:hamilt-ts-rwa}), is expected to lose accuracy as Eq. (\ref{casesr}) would imply in $\Omega=0$.
Nevertheless, it was able to keep $\mathcal{F} > 98\%$ throughout the period. 
In Fig. \ref{fig:fid-slow2} we set $\Omega = 0.1 \nu$, which is already a bit too strong to assure the validity of the RWA leading to red sideband Hamiltonian, as it can be seen.
Interestingly enough, our result still managed to keep $\mathcal{F} > 98\%$, even though it was further from exact resonance $\wt = \nu$.

\section{Final remarks}\label{remarks}

To summarize, we were able to theoretically expand the usage of the trapped ion setup beyond sideband Hamiltonians in two fronts. Firstly, we adapted a protocol of reconstruction of the motional state for the intermediate intensity regime, $\Omega = \nu$, allowing for a speed gain of a least a factor of $10$. The numerical results presented supported the advantage of this result in the presence of a noise source, modeled here as a dephasing channel acting on the electronic subspace of the trapped ion.

Secondly, we were able to extend the known effective Hamiltonian for the intermediate intensity regime, $\Omega \sim \nu$, in the case of a laser that is not in exact resonance with the electronic transition frequency, $\delta \neq 0$. Moreover, this new result presented itself as a bridge between low and intermediate intensity regimes, functioning as a good approximation for the trapped ion Hamiltonian in the whole range $0 < \Omega \leq \nu$, provided we adjust $\delta$ accordingly.

Furthermore, we hope that the results presented here, which expand on the theoretical description of this system in regimes not yet fully explored, will boost research on new protocols and quantum operations in this important setup of laser driven trapped ions.


\section*{ACKNOWLEDGEMENTS} 
T. T. acknowledges financial  support from Coordena\c{c}\~ao de Aperfei\c{c}oamento de Pessoal de N\'ivel Superior (CAPES).
F. L. S. acknowledges partial support from  Funda\c c\~ao de Amparo a Pesquisa do Estado de S\~ao Paulo (FAPESP) Process No. 2021/14135-1, Brazilian National Institute of Science and Technology of Quantum Information (CNPq-INCT-IQ 465469/2014-0), Conselho Nacional de Desenvolvimento Cient\'ifico e Tecnol\'ogico (CNPq) Grant No. 302900/2017-9, and CAPES/PrInt – Process No. 88881.310346/2018-01. We would like to thank Frederico Brito for useful discussions.


\begin{thebibliography}{99}

\bibitem{review} D. Leibfried, R. Blatt, C. Monroe, and D. Wineland, Rev. Mod. Phys. \textbf{75}, 281 (2003).

\bibitem{wine} D. Leibfried, D. M. Meekhof, B. E. King, C. Monroe, W. M. Itano and D. J. Wineland, Phys. Rev. Lett. \textbf{77}, 4281 (1996).

\bibitem{matthias}  H. Takahashi, E. Kassa, C. Christoforou and M. Keller, Phys. Rev. Lett. \textbf{124}, 013602 (2020).

\bibitem{gato}  F. L. Semi\~ao, A. Vidiella-Barranco, Phys. Rev. A \textbf{71}, 065802 (2005). 

\bibitem{thermali} G Kirchmair et al., New J. Phys. \textbf{11}, 023002 (2009).

\bibitem{Hastrup} J. Hastrup et al., Phys. Rev. Lett. \textbf{126}, 153602 (2021).

\bibitem{cz} J. I. Cirac , P. Zoller, Phys. Rev. Lett. \textbf{74}, 4091 (1995).

\bibitem{qre} J. F. Poyatos, J. I. Cirac, and P. Zoller, Phys. Rev. Lett. \textbf{77}, 4728 (1996).

\bibitem{Leung} P. H. Leung et al., Phys. Rev. Lett. \textbf{120}, 020501 (2018).

\bibitem{qreo}  W. S. Teixeira, M. K. Keller, and F. L. Semi\~ao, New J. Phys. \textbf{24}, 023027 (2022). 

\bibitem{rwaperturb} F. Nicacio and F. L. Semi\~ao, J. Phys. A: Math. Theor. \textbf{49}, 375303 (2016).

\bibitem{Jonathan} D. Jonathan, M. B. Plenio, and P. L. Knight, Phys. Rev. A \textbf{61}, 042307 (2000).

\bibitem{Cessa}  H. Moya-Cessa, A. Vidiella-Barranco, J. A. Roversi, D. S. Freitas, and S. M. Dutra, Phys. Rev. A \textbf{59}, 2518 (1999).

\bibitem{rabi0} M. -J. Hwang, R. Puebla, and M. B. Plenio, Phys. Rev. Lett. \textbf{115}, 180404 (2015).

\bibitem{rabi1} R. Puebla, M. -J. Hwang, and M. B. Plenio, Phys. Rev. A \textbf{94}, 023835 (2016).

\bibitem{wigner1}  D. Leibfried et al., Phys. Rev. Lett. \textbf{77}, 4281 (1996).

\bibitem{wigner2} L. G. Lutterbach and L. Davidovich, Phys. Rev. Lett. \textbf{78}, 2547 (1997).

\bibitem{interfer} J. F. Poyatos, J. I. Cirac, R. Blatt, and P. Zoller, Phys. Rev. A \textbf{54}, 1532 (1996).

\bibitem{cessa-review} H. Moya-Cessa, F. Soto-Eguibar, J. M. Vargas-Mart\'inez, R. Ju\'arez-Amaro, and A. Z\'u\~niga-Segundo, Phys. Rep. \tb{513}, 229 (2012).

\bibitem{cahilglauber} K. E. Cahill and R. J. Glauber, Phys. Rev. \textbf{177}, 1882 (1969).

\bibitem{decoh1} D. J. Wineland, C. Monroe, W. M. Itano, D. Leibfried, B. E. King, and D. M. Meekhof, J. Res. Natl. Inst. Stand. Technol. \textbf{103}, 259 (1998).

\bibitem{decoh2} S. Schneider and G. J. Milburn, Phys. Rev. A \tb{57}, 3748 (1998).

\bibitem{decoh3} C. Di Fidio and W. Vogel, Phys. Rev. A \tb{62}, 031802(R) (2000).

\bibitem{decoh4} S. Brouard and J. Plata, Phys. Rev. A \tb{70}, 013413 (2004).

\bibitem{decoh5} R. Ozeri, W. M. Itano, R. B. Blakestad, J. Britton, J. Chiaverini, J. D. Jost, C. Langer, D. Leibfried, R. Reichle, S. Seidelin, J. H. Wesenberg, and D. J. Wineland, Phys. Rev. A \tb{75}, 042329 (2007).

\bibitem{decoh6} R. Bonifacio, S. Olivares, P. Tombesi, and D. Vitali, J. Mod. Opt. \tb{47}, 2199 (2000).

\bibitem{qojulia} S. Krämer, D. Plankensteiner, L. Ostermann and H. Ritsch., Comp. Phys. Comm. \tb{227}, 109 (2018).

\bibitem{julialang} J. Bezanson, A. Edelman, S. Karpinski and V. B. Shah., SIAM Rev. Soc. Ind. Appl. Math. \tb{59}, 65 (2017).

\bibitem{fabry-perot} P. Kobel, M. Breyer, and K\"ohl, npj Quantum Inf. \tb{7}, 6 (2021).

\end{thebibliography}
\end{document}